# Temporal pattern recognition with delayed feedback spin-torque nano-oscillators

M. Riou,[1] J. Torrejon,[1] B. Garitaine,[1] F. Abreu Araujo,[1] P. Bortolotti[1], V. Cros,[1] S. Tsunegi,[3] K. Yakushiji,[3] A. Fukushima,[3] H. Kubota,[3] S. Yuasa,[3] D. Querlioz,[4] M. D. Stiles,[2] J. Grollier[1]

[1]*Unité Mixte de Physique CNRS, Thales,Université Paris-Sud, Université Paris-Saclay, 91767 Palaiseau, France*

[2]*Center for Nanoscale Science and Technology, National Institute of Standards and Technology, Gaithersburg, Maryland 20899-6202, USA.*

[3]*National Institute of Advanced Industrial Science and Technology (AIST), Spintronic Research Center, Tsukuba, Ibaraki 305-8568, Japan*

[4]*Center for Nanoscience and Nanotechnology, CNRS, Université Paris-Sud, Université Paris-Saclay, 91405, Orsay, France*

The recent demonstration of neuromorphic computing with spin-torque nano-oscillators has opened a path to energy efficient data processing. The success of this demonstration hinged on the intrinsic short-term memory of the oscillators. In this study, we extend the memory of the spin-torque nano-oscillators through time-delayed feedback. We leverage this extrinsic memory to increase the efficiency of solving pattern recognition tasks that require memory to discriminate different inputs. The large tunability of these non-linear oscillators allows us to control and optimize the delayed feedback memory using different operating conditions of applied current and magnetic field.

## I. INTRODUCTION

Recurrence can play a powerful role in information processing. It is thought to provide a source of memory in the brain and allows recurrent artificial neural networks to process sequences [1], [2]. When a network is recurrent, input data can remain present in the network for an extended time creating dynamical memory. The state of the network not only depends on the current input value, but also on past values. For neuromorphic computing, recurrence has been adapted in hardware by combining a feedback loop with the physical component that plays the role of the neuron. This approach has been tested in different systems (electronic, optical, photonic) for tasks that require memory such as chaotic series prediction [3]–[11].

Here, we implement recurrence in lower power nano-devices called spin-torque nano-oscillators, [12], [13] which are composed of two ferromagnetic layers separated by a non-magnetic material. In these devices, magnetization dynamics is driven by spin-polarized current through an effect known as spin-transfer torque. When current flows through the magnetic multilayer, the current becomes spin polarized and transfers angular momentum between the magnetic layers, resulting in a torque on the magnetization. For high enough current density, this torque can destabilize the magnetization of the free layer, which is then set in a sustained precessional state. These magnetic oscillations are converted into resistance oscillations due to magnetoresistance effects: the device then functions as an electrical nano-oscillator. The frequency of these oscillators can be tuned from a few hundreds of megahertz to tens of gigahertz by varying the applied electrical current and magnetic field.

This control makes spin-torque nano-oscillators promising as nanoscale microwave sources. Moreover these oscillators have the all essential features needed as hardware neurons because they have nanometer size, they can interact through their emitted magnetic fields [14], [15] or electrical currents [16] to emulate synaptic coupling, the amplitude of the oscillation depends non-linearly on the input current, and they have an intrinsic memory related to the relaxation of magnetization. Using these features, neuromorphic computing with spin-torque oscillators was recently demonstrated [17]–[20]. However the intrinsic memory of these oscillators derived solely from their relaxation time, which is quite short: tens to hundreds of nanoseconds for vortex gyration [17], [18] and even less for other types of magneto-dynamical modes such as uniform precession, bullet modes, or spin waves. This short time limits the number of cognitive tasks that can be performed. In particular, sequence problems such as chaotic series prediction or natural language processing require longer memory to recognize temporal patterns. Delayed feedback spin torque oscillators have been studied both theoretically [21], [22] and experimentally [23], [24]. But the delay in these studies was much smaller (tens of nanoseconds) and applications were mainly related to the reduction of phase noise as well as the optimization of emitted power and linewidth.

In this study, we use a delayed-feedback spin-torque nano-oscillator to remember and recognize patterns. In Sec. II, we show that adding delayed-feedback to the oscillator can enhance the range of the memory from one relaxation time (≈ 200 ns) to tens of relaxation times. In Sec. III, we emulate a reservoir computer with a single, time-multiplexed oscillator [25], [26], and demonstrate the performance of the extrinsic memory by recognizing temporal patterns composed of random sine and square waveforms. Finally, in Sec. IV, we vary the current and magnetic field applied to the nano-oscillator to find the optimal operating regime of the oscillator for pattern recognition.

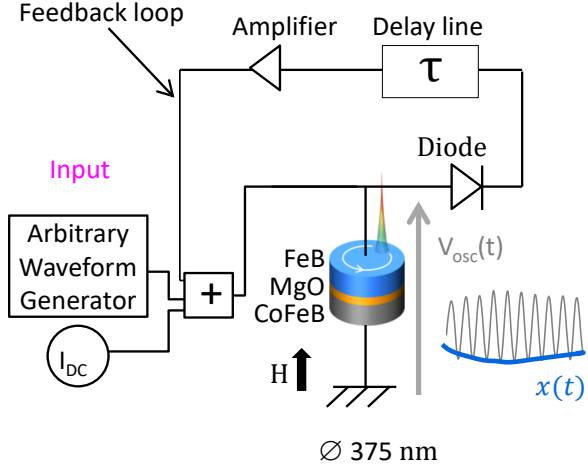

FIG 1: Experimental set up. The spin-torque oscillator is subjected to DC current and perpendicular magnetic field which set the operating point. It emits an oscillating voltage $V_{osc}(t)$. The time-varying input is generated by an arbitrary waveform generator. A diode allows measuring directly the amplitude of the oscillations $x(t)$, which is used for computation. The feedback loop consists of an electronic delay line ($\tau$ =4.3 μs) and an amplifier (the total amplification in the line is +20 dB). The signals are added with power splitters.

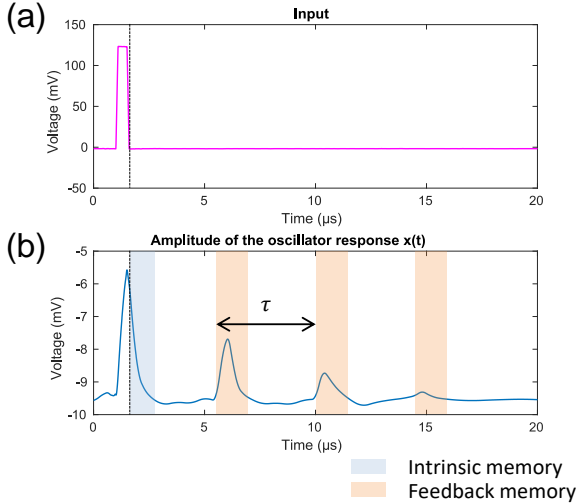

FIG 2: (a) Input spike (in magenta) sent by the arbitrary waveform generator to the spin-torque oscillator. (b) Blue curve: variation in the amplitude of the delayed feedback oscillator response $x(t)$. The shaded areas in blue and orange indicate the intrinsic and feedback memory respectively. The operating point is 600 mT and -6.5 mA.

## II. EXPERIMENTAL IMPLEMENTATION OF DELAYED FEEDBACK SPIN TORQUE OSCILLATOR

The experimental set up is illustrated in Fig 1. The nano-oscillator is a cylindrical magnetic tunnel junction of diameter 375 nm. The pinned layer is a CoFeB-based synthetic antiferromagnet, the tunnel barrier is MgO and the free layer is a 3 nm thick FeB layer. For this aspect ratio of the free layer, the vortex magnetization is the stable configuration in the ground state. Such oscillators exhibit a high power (a few microwatts) over noise ratio ($\approx 100$), low phase noise [27], and show optimal microwave properties for neuromorphic computing [17]–[20]. The oscillator is connected to a DC current source and is subjected to an external magnetic field to induce gyrotropic motion of vortex core via spin transfer torque mechanism. These two tunable parameters set the regime where the oscillator operates.

The classification process used in the next section starts by encoding the different patterns in time-varying electrical input signals generated by an arbitrary waveform generator. This electrical input signal moves the oscillator out of its operating point, thus changing the amplitude and phase of the oscillator voltage response $V_{osc}(t)$ (in grey in Fig. 1). For simplicity, only the amplitude $x(t)$ (in blue in Fig. 1(a)) of the oscillations is used for computing. The amplitude is measured by placing a diode after the oscillator to capture the envelope of the oscillating signal. The delayed feedback loop is composed of an electronic delay line with a delay $\tau = 4.3$ µs and a total amplification of about 20 dB. The addition of the delayed feedback to the input is made with a power splitter. The emitted voltage of the oscillator (10 mV to 15 mV) is reinjected by the delay line. The reinjected voltage level depends on the signal amplitude emitted by the oscillator itself $x(t)$ and thus depends on the operating point (magnetic field and current). For the highest emitted voltage (at low magnetic field), the reinjected amplitude can reach approximately 50 % of the amplitude of the input signal (250 mV peak to peak). Finally, the oscillation amplitude is recorded with an oscilloscope.

The intrinsic memory of the oscillator and the feedback memory are illustrated in Fig. 2. A current pulse of 200 ns duration is sent to the oscillator (in magenta in Fig. 2(a)) and the voltage amplitude of oscillator $x(t)$ is recorded by an oscilloscope (in blue in Fig. 2(b)) during a much longer time (20 µs) to observe the reinjection effects and the memory induced by the delay line. The input spike creates a perturbation in the oscillator and modifies the orbit of gyrotropic motion of the vortex core. This changes the amplitude of the signal emitted by the oscillator $x(t)$. After the spike, the vortex core returns to the orbit defined by the fixed magnetic field and DC current: the amplitude of the oscillations $x(t)$ returns progressively to its initial level. During this time, which corresponds to the relaxation, the oscillator still remembers the input because the oscillation amplitude has not returned to its initial level. The relaxation time of magnetization dynamics, roughly defined by the magnetic damping and the frequency of oscillator ($T_{relax} \sim \frac{1}{\alpha f}$), is around 200 ns, which corresponds to the range of the intrinsic memory of the oscillator (except when the oscillator operates in the immediate proximity of the critical current, where the relaxation time increases but the emission level is very low). This intrinsic memory, highlighted in blue in Fig. 2(b), has a duration of a few hundreds of nanoseconds.

The feedback is implemented by propagating the perturbation of the oscillation amplitude in the delay line for $\tau = 4.3$ µs and reinjecting it into the oscillator. This injection induces new variations in the amplitude of the oscillator response $x(t)$. Indeed, echoes in the oscillator response are observed every τ after the end of the input signal in Fig. 1(c). These echoes are the manifestation of the external memory provided by the delayed feedback. Since after each τ the echo is more and more attenuated, the feedback memory is a fading memory. The fact that memory fades is important to process temporal sequence for which only the recent history is important [3]. For a magnetic field of 600 mT and a DC current of -6.5 mA, we observe variations of the oscillator output even after 13 μs (approximately 60 relaxation times). The range of the feedback memory depends of the delay in the line. Here we observe a memory of 13 μs, which corresponds to three times the delay of the line, but choosing a longer delay in the line would have resulted in a longer range of memory. Moreover, the feedback memory is sparse, that is, the information that the oscillator received as input in the past is only accessible every time τ (4.3 μs in our case) in the areas highlighted in orange in Fig. 2(b). In between, this information cannot be extracted from the measured trace. If the input is discrete, the time step of the input and the delay in the line are often chosen to be equal so the oscillator remembers the input at previous time steps.

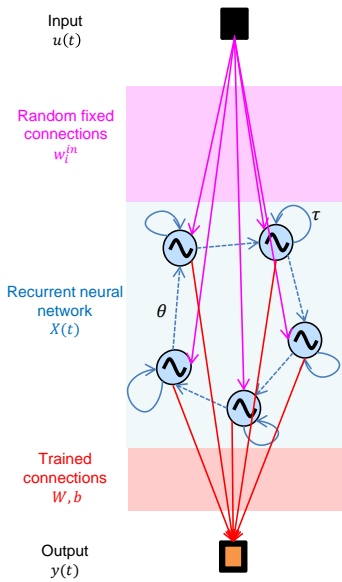

FIG 3: Principle of reservoir computing. The input $u(t)$ is connected via fixed connections $w_i^{in}$ to a recurrent neural network, called a reservoir. The reservoir maps the input $u(t)$ into a higher dimensional state $X(t)$, that is, each neuron output is a coordinate of the projected input. The output $y(t)$ is obtained by combining the neuron responses with trained connections $W, b$.

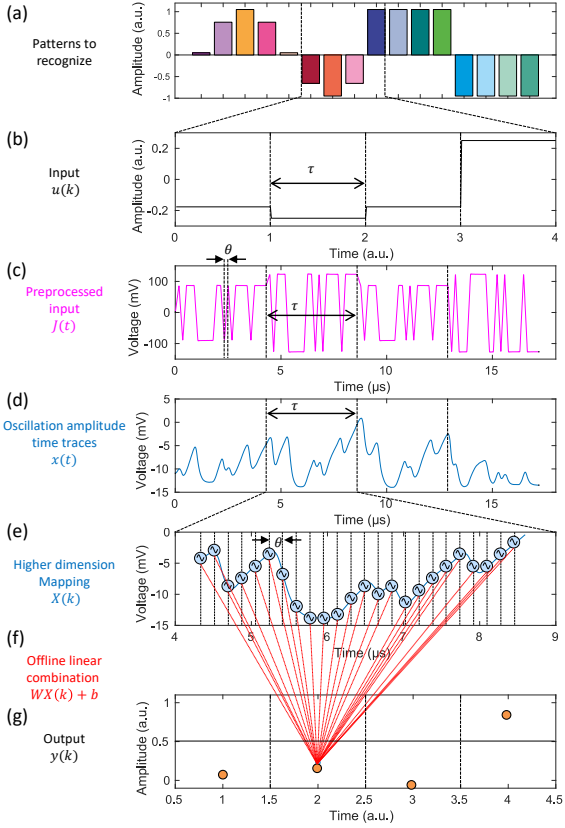

FIG 4: Implementation of reservoir computing with a single oscillator. (a) The patterns to recognize are discretized sine and square periods. (b) The discrete input $u(k)$ is a time sequence of these sine and square periods, randomly arranged. (c) Preprocessed input $J(t)$ obtained by multiplying $u(k)$ with a fast varying sequence called a mask. This mask varies every θ and represents the coupling $w_i^{in}$ between the input and each virtual neuron of the reservoir. (d) $x(t)$ is the oscillator response in oscillation amplitude to the preprocessed input $J(t)$. The oscillator plays the role of all virtual neurons one after the other, every θ. (e) The higher dimension mapping $X(k)$ of the input $u(k)$ can be retrieved in the different sequences of $x(t)$ of duration τ. In our experiment the oscillator emulates $n = 24$ neurons, $\theta = 180$ ns, $\tau = 4.3$ μs. (f) Then the output $y(k)$ is reconstructed offline on a computer by combining linearly the responses of the virtual neurons $y(k) = W.X(k) + b$. (g) If the output $y(k)$ is less than 0.5, the input $u(k)$ is classified as a sine, otherwise it is classified as a square.

## III.  NEUROMORPHIC COMPUTING APPROACH: RESERVOIR COMPUTING

To test the efficiency of the extrinsic memory induced by reinjection for computation, we perform pattern recognition of sine and square waveforms. To perform this task, we have adapted the delayed feedback spin torque oscillator to a recurrent neural network approach, called reservoir computing, which has been identified to be suitable for different hardware implementations: optical [3], [4], photonic [5], [11] and spintronic devices [17], [18], [20]. Fig. 3 shows the architecture of a general reservoir computing scheme. The input $u(t)$ (in black in Fig. 3) is coupled with fixed connections $w_i^{In}$ (in magenta in Fig. 3) to a recurrent neural network referred to as reservoir (in blue in Fig. 3) where the internal connections between neurons are chosen to be fixed and random. The reservoir maps the input $u(t)$ in a higher dimensional state referred as reservoir state $X(t)$, where the output of each neuron represents a transformation of the projected input $u(t)$ (note that the dimension of $X(t)$ is equal to the number of neurons $n$). If the reservoir is complex enough (which means that it is on the one hand of high enough dimension, with sufficient non-linearity and on the other it has sufficient memory), the transformed problem becomes linearly separable in the reservoir state. The output $y(t)$ is obtained by combining the neuron outputs ($x_i(t)$ values) with trained connections $W, b$.

The experimental implementation of reservoir computing with multiple interconnected devices is challenging, but a simplified approach has been successfully demonstrated [3] based on constructing a reservoir from a time-multiplexed single device. Here, the entire recurrent network is replaced by a single non-linear node which serves as each virtual neuron one after the other. In our case, the non-linear node is a spin-torque oscillator. The input is projected to higher dimension in time instead of in multiple devices (see figures Fig. 4(a-g)). The input stream $u$ is an aggregation of discretized sine and square periods (patterns in Fig. 4(a)). The goal is to return an output 0 if the input value $u(k)$ belongs to a sine and an output of 1 if the input value $u(k)$ belongs to a square. The system processes a discretized version of the input stream $u(t)$ with each input value $u(k)$ (in black in Fig. 4(b)) being applied $n$ times into the reservoir (where $n$ is the number of virtual neurons) with a time step $\theta$. The varying connections between the input and the virtual nodes of the reservoir are captured by multiplying $u(k)$ by a sequence that varies every time step $\theta$ and repeats every $\tau = n\theta$, the time between different input values. We choose the connection between the input and the virtual neurons to be either +1 or -1. The resulting preprocessed input $J(t)$ (see Fig. 4(c)) is applied to the oscillator.

The oscillator then emits a transient amplitude response $x(t)$ (Fig. 4(d)). The responses of the $n$ neurons $x_i(t)$ in a spatial reservoir are here replaced by the response of the single oscillator $x(t)$ sampled $n$ times over intervals of length $\theta$ during a time $\tau$ (Fig. 4(e)). If the time $\theta$ is shorter than the relaxation time, $x(t)$ depends on $x(t - \theta)$. This situation is analogous to a neural network consisting of multiple connected devices. Here the coupling between the devices is provided by the influence of previous states of the oscillator on its current state. If a delayed feedback loop with a delay $\tau = n\theta$ is added to the oscillator, the oscillator also receives the signal $x(t - \tau)$ at time $t$. The virtual neurons $x_i$ therefore receive their own past output, so they are recurrently coupled to themselves. The equivalent multiple-device neuron network architecture is represented in blue in Fig. 3. Dashed arrows are the connections emulated by the relaxation and solid line feedback arrows are the connections emulated by the delayed feedback loop. $X(k)$ (dimension (n,1)) represents the mapping $u(k)$ into a higher dimensional space that is obtained from discrete values sampled every $\theta$ in each interval of length $\tau = n\theta$ in the time trace $x(t)$ (highlighted by circles in Fig. 3(d)). The output value is obtained by the linear combination $y(k) = W \cdot X(k) + b$ (Fig. 4(f)) (with $W$ of dimension $(1, n)$). If the output $y(k)$ is less than 0.5, the input $u(k)$ is classified as a sine. Otherwise, it is classified as a square (Fig. 4(g)).

In our experiment, we use $n = 24$ virtual neurons. The oscillator acts as each virtual neuron every

$\theta = 180$ ns time interval. To emphasize the role of the memory effect brought by delayed feedback, we choose a larger time $\theta$ (comparable with the relaxation time) than that used in the previous spintronic implementation without delayed feedback [17], [18]. The longer interval attenuates the intrinsic oscillator memory effect so as to focus on the memory provided by the feedback. The input $u(k)$ is processed with a time step $\tau = 4.3$ µs, which corresponds to the delay time in the feedback circuit. The preprocessed input $J(t)$ is created by the arbitrary waveform generator. The oscillator amplitude response $x(t)$ is recorded with an oscilloscope and the higher dimension mappings $X(k)$ are extracted from $x(t)$. The optimal output weights $\mathbf{W}$, $b$, which are determined during the learning phase, and the output $y(k)$ are computed offline on a computer.

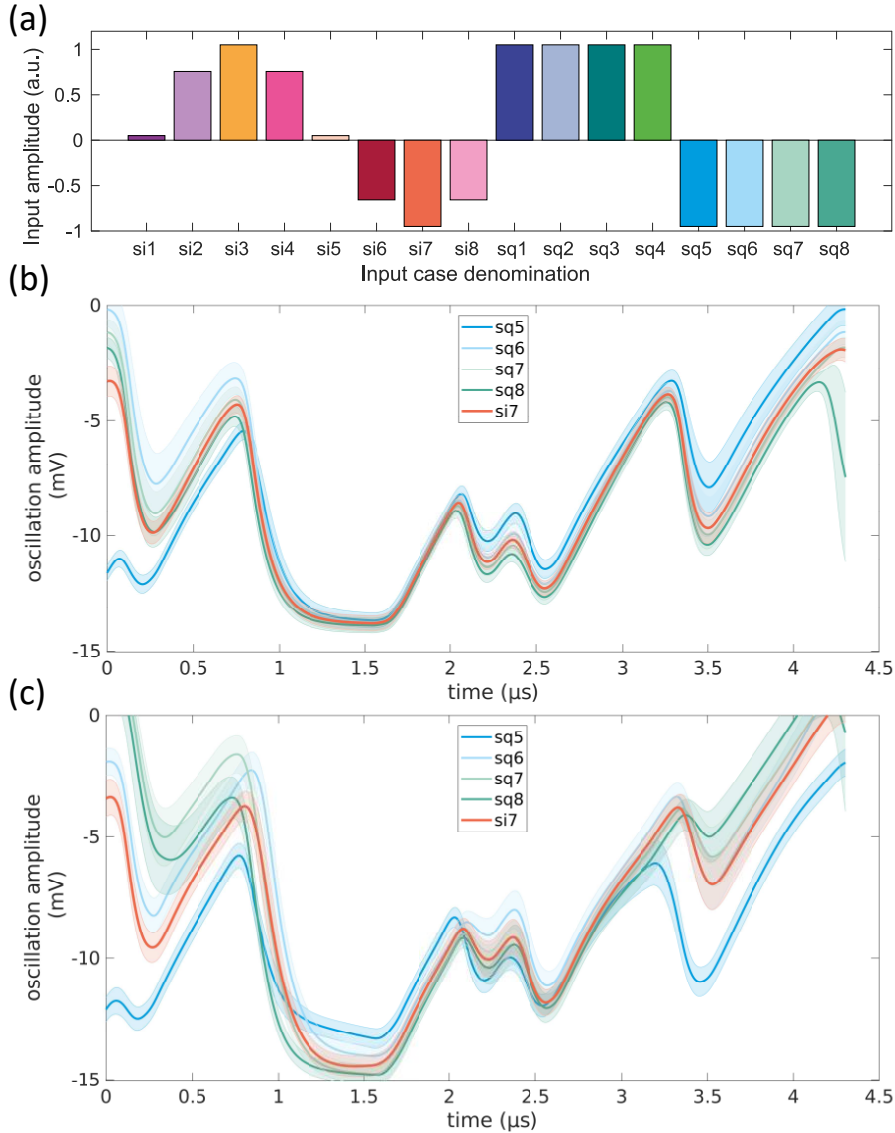

FIG 5 : (a) Pattern to recognize: the time-varying input $\mathbf{u}(\mathbf{k})$ used to evaluate the delayed feedback spin torque oscillator is composed of sine and square sequences. The 8 inputs from sine are designated as si1-8 and the 8 inputs of a square are designated as sq1-8. (b) Comparison of the average experimental time traces of $\mathbf{x}(\mathbf{t})$ without feedback for si7 (represented in orange) and sq5-8

(represented in cool colors, blues and greens). The shaded regions indicate twice the standard deviation of $x(t)$ for sq5, sq8, and si7. The operating point is 300mT and -6.5mA. (c) Similar comparison with feedback.

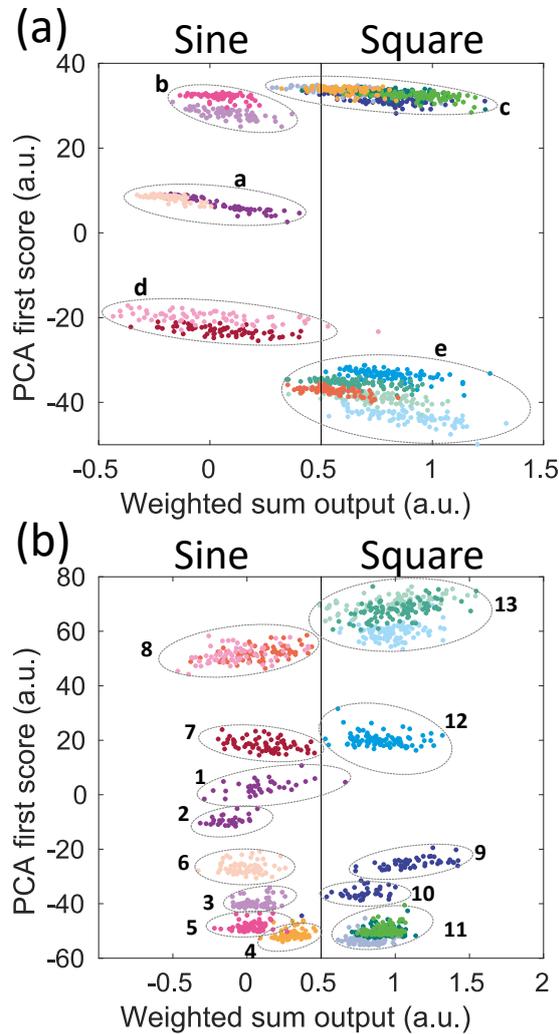

FIG 6: (a) 2D visualization of the higher dimension mapping $\mathbf{X}(\mathbf{k})$ obtained from experiment for all the different inputs $\mathbf{u}(\mathbf{k})$ without feedback. The operating point is 300 mT and -6.5 mA. The separation between sine and square cases is the vertical black solid x = 0.5. For a perfect recognition, points corresponding to sine input should be left to this line and points corresponding to square input should be to the right. Without feedback, the points for si7 (orange) and si3 (yellow) are mixed up with the corresponding square cases. The error rate is 10.8 % in the test phase (69 errors). Five clusters are observed corresponding to the five different input values. These clusters are highlighted by grey ellipses and denominated by the letters a to e. (b) 2D visualization of the higher dimension mapping $\mathbf{X}(\mathbf{k})$ obtained from experiment for all the different inputs $\mathbf{u}(\mathbf{k})$ with feedback. The operating point is 300 mT and -6.5 mA. Thirteen clusters are observed confirming that feedback enables the oscillator to separate inputs by remembering sequences of two consecutive inputs. These clusters highlighted by grey ellipses are denominated by the numbers 1 to 13. The points for si7 and si3 are different from the square cases and fall to the left of the vertical separation line. The final classification is almost perfect with only 0.16 % error rate (one error) during the test phase.

## IV. PATTERN RECOGNITION USING DELAYED FEEDBACK SPIN TORQUE NANO-OSCILLATOR

We choose to classify sine and square waveforms, as done in previous studies [4], [10], [17], [18], to evaluate the performance of our delayed-feedback reservoir computing based on a spin torque oscillator. Each period of the waveforms is discretized in 8 points giving 16 different inputs (Fig. 5(a)). We refer to the 8 points of the sine waveform as si1-8, which are represented in warm colors, and the 8 points of the square waveform as sq1-8, which are represented in cool colors in Fig. 5(a). Because of degeneracies, the inputs take 5 different values in a sine waveform and two different values in a square waveform. To return the same output value for 5 different input values of a sine (or 2 in the case of the square), the reservoir must be non-linear. In the sine period, the $3^{rd}$ and the $7^{th}$ point (referred as si3 and si7 see Fig. 5(a)) have the values +1 and -1, which are the same values as in the square waveform. In the absence of memory, it is impossible to classify the input value of +1 or -1 as belonging either to a sine or to square waveform. Therefore, this temporal pattern recognition task needs both non-linearity and memory in the neural network. The input $u(k)$ is composed of 1280 points (160 periods of sine or square randomly arranged). The first half of the points is used for training (to find optimum output weights $W$ and $b$) and the second half for testing. The operating point of the oscillator is 300 mT and -6.5 mA.

Figures 5(b) and 5(c) show the average $x(t)$ response over all the cases where input is the same and twice the standard deviation for the si7, sq5, and sq8 cases with and without feedback. First, without feedback, we can see that time traces $x(t)$ for si7 overlap within the two-standard-deviation uncertainties with time traces $x(t)$ for sq5 to sq8 (Fig. 5(b)). When the feedback memory is included (Fig. 5(c)), the time trace si7 differs from the respective sq time traces. The feedback also breaks the symmetry between the sq points with the same input values. The traces for sq5 becomes significantly different from the traces for sq6 to sq8 (see Fig. 5(c)).

To analyze the performance of the oscillator in the absence or presence of delayed feedback, we plot functions of $X(k)$ in a 2D map in Fig. 6(a) and (b). As mentioned in the previous section, each $\tau$ long time trace corresponds to the mapping in 24 dimensions $X(k)$ of an input value $u(k)$. To visualize the separation, the mappings $X(k)$ are projected linearly in two dimensions in (Fig. 6(a) and (b)). To see the separation between the sine and the square, the first coordinate (x axis) of the 2D representation is given by the linear combination $W \cdot X + b$. It can be seen geometrically as a projection of the data along the weight vector $W$. In this 2D projection, the separation between sine and square is defined by the vertical line at $x = 0.5$ in Fig 6(a-b). Ideally the sine cases (si1-8) should all fall on the left of this line and square cases (sq1-8) on the right, respectively.

The second coordinate (y axis) of the projected data points is the first component from a principal component analysis (PCA) of the data reduced in the space orthogonal to $W$. The PCA separates the clusters (collections of points that are neighbors in the 24-dimension reservoir state) in the data. The

PCA is performed in the space orthogonal to **W** to project the data along two orthogonal vectors. Without feedback, 5 clusters denominated a to e are observed in the 2D map. They correspond respectively to the 5 different values taken by the input: 0 (si1 and si5), 0.71 (si2 and si4), 1 (si3 and sq1-sq4), -0.71 (si6 and si8) and -1 (si7 and sq5-sq8). The ambiguous cases corresponding to si3 and si7 are completely mixed up with sq1-4 and sq5-8, respectively. Since the time step $\theta$ has been chosen close to the relaxation time of the oscillator, the intrinsic memory is too small to remember a two-point sequence. With no feedback, the error rate for the identification of sine and square is 10.8 % for this value of $\theta$. All the 69 errors during test phase are due to bad classification between si3 and sq1-4 or si7 and sq5-8. These results are in excellent agreement with the prediction and the qualitative analysis of the time traces.

Without memory, we expect five clusters, because the input takes only five different values. With memory, new clusters appear and the number of clusters depends of the range of the memory. If the reservoir remembers one time step, the clusters in the higher dimension mapping should correspond to the different sequences of two consecutive input points at time k and k-1. We expect nine different clusters for the sine and five different clusters for the square. Similarly, if the reservoir can remember two time steps in the past, eighteen clusters are expected.

With delayed feedback to provide memory, the 2D map shows 13 clusters instead of the 5 found without delayed feedback. This symmetry breaking can be explained by a memory of one time step in the past. We observed 13 (and not 14 clusters as expected) on the 2D projection because points for si7 and si8 overlap (cluster 8 in Fig. 6(b)). The first component of the PCA may have not conserved all the clusters present in the initial 24-dimension reservoir state. However, all the other 13 expected clusters are clearly separated in the 2D projection, confirming that the delayed feedback spin torque nano-oscillator remembers sequence of two consecutive inputs. With delayed feedback, si3 and si7 become different from the square traces and a linear separation can be found with a very small error rate (0.16 % errors on the test set which corresponds to only one error). Adding delayed feedback to the oscillator allows it to distinguish si3 and si7 from square points because the feedback adds a memory of the previous input point. For the optimal operating point (300 mT and -6.5 mA), the feedback is very efficient and suppresses 98.6 % of the errors (Fig 6(a-b), Fig 9), demonstrating the efficiency of the feedback memory for computation.

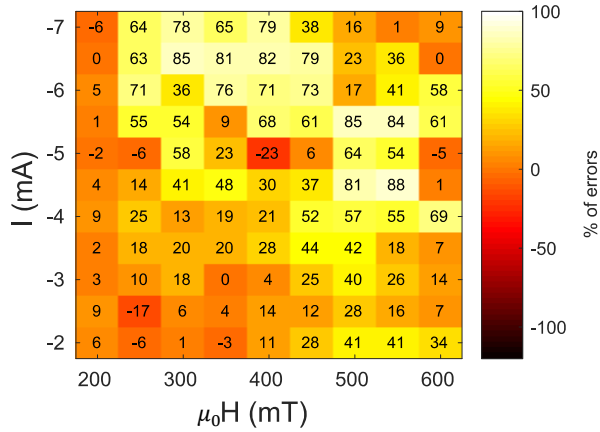

FIG 7: Reduction of errors on the ambiguous cases si3 and si7 due to feedback depending of the operating point, Eq. (1): this reduction of the error is evaluated on the training set. The number of suppressed errors is renormalized by the total number of errors on the training set without feedback. Brighter colors indicate high reduction of the error. The magnetic field is swept from 200 mT to 600 mT and the DC current is swept from -2 mA to -7 mA. In 90 % of operating points, feedback helps distinguishing si3 and si7 during the training phase. Error reduction is high for high DC current (larger than -4 mA) and magnetic field between 250 mT and 400 mT

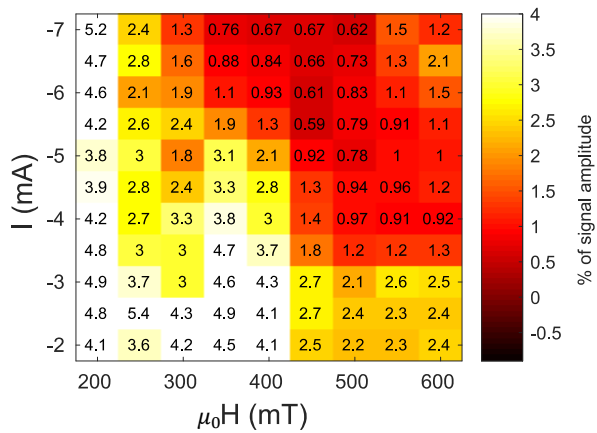

FIG 8: Normalized noise level as a function of the operating point: the magnetic field is swept from 200 mT to 600 mT and the DC current is swept from -2 mA to -7 mA. At low field (200 mT to 400 mT) and low current (-2 mA to -5 mA) the noise level in the oscillation amplitude response is high. Areas of high noise (brighter color) correspond to areas where the feedback is not efficient at suppressing errors.

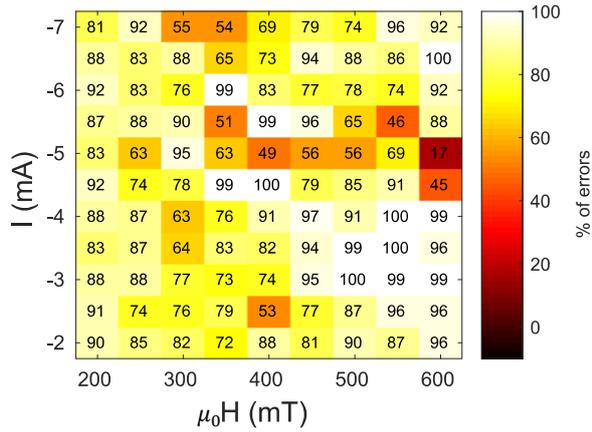

FIG 9: Proportion of si3 and si7 misclassified on training set without feedback, Eq. (2): the magnetic field is swept from 200 mT to 600 mT and the DC current is swept from -2 mA to -7 mA. For 600 mT and -5 mA, si3 and si7 are well classified even without feedback (only 17 % error). Other operating points exhibit good performance (orange and red areas).

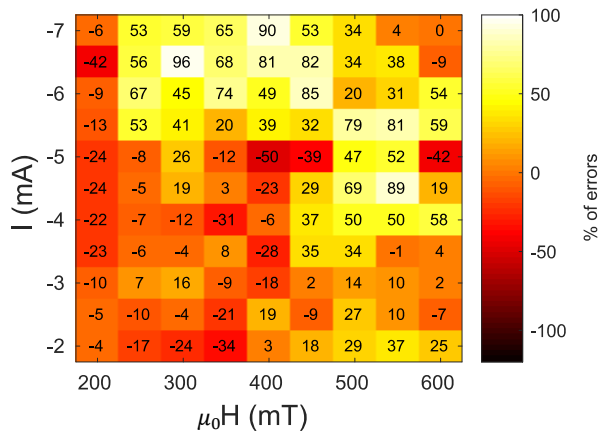

FIG 10: Global error reduction on the training set due to the feedback as a function of the operating point, Eq. (3): the number of suppressed errors is renormalized by the total number of errors on the training set without feedback. The magnetic field is swept from 200 mT to 600 mT and the DC current is swept from -2 mA to -7 mA. Feedback reduces the error rate on the training set only in 66 % of the cases. The feedback thus generates new errors in at least 24 % of the bias points. Brighter colors indicate high reduction of the error. Error reduction is high for high DC current (larger than -4mA) and fields between 250 mT and 400 mT.

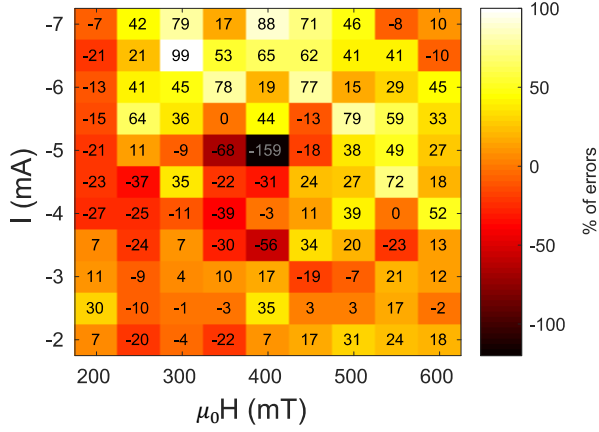

FIG 11: Global error rate reduction due to feedback on the test set as a function of the operating point, Eq. (4): the number of suppressed errors is renormalized by the total number of errors on the test set without feedback. The magnetic field is swept from 200 mT to 600 mT and the DC current is swept from -2 mA to -7 mA. Global error rate reduction during test phase is in good agreement with the improvement during training phase. Feedback improves the result in 60 % of the cases during the test phase. Brighter colors indicate high reduction of the error. Error reduction is high for high DC current (larger than -4 mA) and fields between 250 mT and 400 mT.

Both the magnetic field and the DC current change the non-linear dependence of the voltage oscillation amplitude with the input current (see figure 3 in Ref. 1) with the DC current acting like an offset for the input current. The operating point fixed by these two parameters changes the effectiveness of the feedback memory. As seen previously the reduction of the errors classifying the si3 and si7 cases shows that the memory of the oscillator was enhanced. This improvement is quantified by the proportion

$$Pr_1 = \frac{\left(E_{si3}^{no\ fb}+E_{si7}^{no\ fb}\right)-\left(E_{si3}^{fb}+E_{si7}^{fb}\right)}{2(N_{si3}^{train}+N_{si7}^{train})}, \quad (1)$$

, where $E_{si3}^{no\ fb}$ and $E_{si7}^{no\ fb}$ are respectively the number of misclassified si3 and si7 examples during training in the case without feedback, $E_{si3}^{fb}$ and $E_{si7}^{fb}$ are respectively the number of misclassified si3 and si7 examples during training phase in the case with feedback and $N_{si3}^{train}$ and $N_{si7}^{train}$ are the total number of si3 and si7 examples during training phase. Fig. 7 is a 2D map of this quantity as a function of the operating point, where bright colors indicate high improvements and dark colors low improvement or, in the case where $Pr_1$ is negative, that more cases si3 and si7 are misclassified with feedback. In 90% of the operating point conditions the feedback improves the recognition of the si3 and si7 cases (see Fig. 7), showing the memory it brings breaks the degeneracies between inputs from square and sine periods. The low improvements are mainly at low magnetic fields: 200 mT and between 250 mT and 400 mT and between -2.0 mA to -5.0 mA. This area correlates well with regions where the amplitude noise of the oscillator is high compared to the signal amplitude (see Fig 8, bright region).

Some other cases where $Pr_1$ is negative or positive but low are observed on isolated operating points such as 600 mT and -4.5 mA, 600 mT and -5.0 mA, 400 mT and -5.0 mA, 450 mT and -5.0 mA. Fig. 9 shows

$$Pr_2 = \frac{\left(E_{si3}^{no\,fb}+E_{si7}^{no\,fb}\right)}{\left(N_{si3}^{train}+N_{si7}^{train}\right)}, \quad (2)$$

which is the proportion of si3 and si7 inputs misclassified during the training phase in the case without feedback. For these operating points, where higher relaxation time are measured, using the oscillator without feedback already leads to partly good classification of the si3 and si7 cases (see Fig. 9). Indeed with a 180 ns θ time step, the intrinsic memory is not completely negligible close to these operating points, resulting in good classification even without feedback. In particular, at 600 mT and -5.0 mA, 87% of the si3 and si7 cases are well classified without feedback. For 400 mT and -5.0 mA, the feedback generates new mistakes on the si3 and si7 cases. In this particular case the feedback works against the intrinsic memory of the oscillator.

The total error improvement for the training phase is computed as
$$\Delta\epsilon_{train} = \frac{\epsilon_{no\,fb}-\epsilon_{fb}}{\epsilon_{no\,fb}}, \quad (3)$$
where $\epsilon_{no\,fb}$ and $\epsilon_{fb}$ are the total error rates respectively without feedback and with feedback, such that $\Delta\epsilon_{train} > 0$ corresponds to an improvement of classification due to the feedback. For the training phase, the feedback improves the result in 66% of the operating point conditions (Fig. 10). Areas where the feedback is detrimental ($\Delta\epsilon_{train} < 0$) correspond to areas where $Pr_1$ was low or negative. If, in the large majority of cases, feedback improves the classification of si3 and si7 cases, it may also generate new errors that in some operating point conditions, overcome the benefit of external memory. These new errors could be due to the increase of dispersion in the reservoir state. Indeed, if feedback suppresses unwanted degeneracies between inputs from square and inputs from sine, it also generates new clusters as seen at the beginning of this section. This effect is reinforced in areas where dispersion is high due to stochastic behavior at threshold current and field for auto-oscillation (for 200 mT and between 250 mT and 400 mT and -2 mA to -5 mA).

Computing the error gain on the testing data

$$\Delta\epsilon_{test} = \frac{\epsilon_{no\,fb}-\epsilon_{fb}}{\epsilon_{no\,fb}}, \tag{4}$$

where $\epsilon_{no\,fb}$ and $\epsilon_{fb}$ are the error rates for the testing data respectively in the case without feedback and with feedback, we found that the dependency with the operating point is similar to $\Delta\epsilon_{train}$ with the feedback improving in 60% of the cases (Fig. 11). Some values of $\Delta\epsilon_{test}$ are worse than $\Delta\epsilon_{train}$ because of bad generalization such as observed for instance at 400 mT and -5.0 mA. The feedback brings memory in a vast majority of the operating point conditions and even though it may bring new types of errors due to an increased dispersion of the data in the reservoir state, it still improves the error rate in a large range of operating point conditions. Best improvements are obtained in areas of low noise when compared to the amplitude and of low intrinsic memory.

**CONCLUSION**

Spin-torque nano-oscillators demonstrate clear memory effects when feedback is added. The feedback oscillator can remember the effect of the input signal tens of times longer than it can just with the intrinsic memory that is defined by its relaxation time. This feedback can be adapted to the time steps of the input by tuning the delay line. We evaluate the efficiency of feedback as a memory using a single spin torque oscillator with feedback as a reservoir computing recurrent network. The single oscillator with time multiplexing projects the initial problem in a higher dimensional state, turning it into a linearly separable problem. We tested this computation scheme on a temporal pattern recognition task, sine and square waveform classification, which requires memory and non-linearity. Choosing the optimal operating point, the error rate drops from 10.8 % without feedback to 0.16% with feedback. Analyzing the different clusters that appear in the data with a 2D projection for the optimal operating point shows that the delayed feedback remembers one time step in the past allowing a nearly perfect separation of the data. We identify the optimal operating point to be when the amplitude of the reinjected level is high and the noise of the oscillator is low (300 mT and -6.5 mA). While we demonstrated the value of adding delayed feedback to a single node reservoir computing scheme, delayed feedback can also be used as memory for a collection of many such oscillators. In this work, we overcome the intrinsic memory limitation of spin-torque oscillators, opening the path to solve complex sequence recognition with networks of spin-torque oscillators.

**Acknowledgements**

This work was supported by the European Research Council ERC under Grant bioSPINspired 682955. F.A.A. is a Research Fellow of the F.R.S.-FNRS.